\documentclass[11pt]{article}

\usepackage{amsmath}
\usepackage{graphicx}
\usepackage{amsfonts}
\usepackage{amssymb}
\usepackage{epsfig}
\usepackage{color}
\usepackage{psfrag}
\usepackage{epstopdf}

\usepackage{caption}
\usepackage{subcaption}

\newcommand{\EQ}{\begin{equation}}
\newcommand{\EN}{\end{equation}}
\newcommand{\be}{\begin{equation}}
\newcommand{\ee}{\end{equation}}
\newcommand{\bea}{\begin{eqnarray}}
\newcommand{\eea}{\end{eqnarray}}

\begin{document} \setcounter{page}{0}
\newpage
\setcounter{page}{0}
\renewcommand{\thefootnote}{\arabic{footnote}}
\newpage
\begin{titlepage}
\begin{flushright}
\end{flushright}
\vspace{0.5cm}
\begin{center}
{\large {\bf On the nature of the spin glass transition}}\\
\vspace{1.8cm}
{\large Gesualdo Delfino}\\
\vspace{0.5cm}
{\em SISSA -- Via Bonomea 265, 34136 Trieste, Italy}\\
{\em INFN sezione di Trieste, 34100 Trieste, Italy}\\
\end{center}
\vspace{1.2cm}

\renewcommand{\thefootnote}{\arabic{footnote}}
\setcounter{footnote}{0}

\begin{abstract}
\noindent
We recently showed that the two-dimensional Ising spin glass allows for a line of renormalization group fixed points which explains properties observed in numerical studies. We observe that this exact result corresponds to enhancement to a one-generator continuous internal symmetry. This finally explains why no finite temperature transition to a spin glass phase is observed in two dimensions. In more than two dimensions, instead, the continuous symmetry can be broken spontaneously and yields a spin glass order parameter which,  for fixed temperature and disorder strength,  takes continuous values in an interval. Such a feature is shared by the order parameter of the known mean field solution of the model with infinite-range interactions, which corresponds to infinitely many dimensions.
\end{abstract}
\end{titlepage}

\newpage
\tableofcontents

\section{Introduction}
Frustrated random magnets are expected to allow a transition to a spin glass phase. The fundamental model is the Ising model with bonds randomly taking both ferromagnetic and antiferromagnetic values \cite{EA}. The mean field solution \cite{Parisi} of the version with infinite-range interactions \cite{SK} effectively corresponds to spatial dimensionality $d=\infty$. In $d=3$ the spin glass phase occurs in the strong disorder regime and has been beyond reach of analytical methods. Added to the fact that numerical studies are severely complicated by the need to take the disorder average on top of the thermal one, this made the nature of the transition and of the spin glass phase the object of a debate which continues since decades (see \cite{BY,KR,Dahlberg} and references therein).

Recently, the first exact access to the spin glass regime in finite dimension has been obtained in $d=2$ \cite{Ising_glass}. This result built on the observation of \cite{random} that the replica method for quenched disorder can be implemented in an exact way at two-dimensional renormalization group fixed points, where the infinitely many generators of conformal symmetry allow to solve the underlying field theory in the scattering framework. Upon generalization of this method from magnetic random fixed points (\cite{colloquium} for a review) to spin glass fixed points, a remarkable result was found \cite{Ising_glass}: the symmetries of the Ising spin glass allow for a line of fixed points, a circumstance exceptional in a physically relevant context and which immediately provides an explanation to features of $d=2$ spin glass criticality observed in numerical studies. 

In the present paper we add an observation with far reaching consequences: the line of fixed points implies enhancement to continuous internal symmetry. Then, the fact that such a symmetry cannot break spontaneously \cite{MW,Hohenberg,Coleman_goldstone} immediately explains the observed lack of finite temperature transition to a spin glass phase in $d=2$. While the line of fixed points is a special property of the two-dimensional case, the continuous internal symmetry that it reveals is present in all dimensions. The spontaneous breaking of this symmetry, which is allowed in $d>2$,  yields a spin glass phase in which the order parameter, for fixed temperature and disorder strength,  takes continuous values in an interval. This also sheds new light on the reason why an order parameter with such a property characterizes Parisi mean field solution in $d=\infty$. 

The paper is organized as follows. In the next section we explain the relation between a line of fixed points in $d=2$ and continuous internal symmetry. The implications for the Ising spin glass are analyzed in section~\ref{Ising_glass}, while the last section contains some final remarks.

\section{Symmetries and criticality}
\label{symmetry}
In field theory a continuous symmetry is associated to a current $J^\mu$ satisfying the conservation equation
\EQ
\partial_\mu J^\mu=0\,,
\label{conservation}
\EN
and to the conserved quantity
\EQ
{\cal Q}=\int d^{d-1}x\,J^0(x)\,.
\label{Q}
\EN
We are concerned with internal symmetries, namely symmetries which do not involve coordinate transformations; they hold along renormalization group trajectories and, in particular, at the fixed points such trajectories emanate from. For such a symmetry, ${\cal Q}$ is dimensionless due to scale invariance at fixed points, and (\ref{Q}) shows that the scaling dimension of $J^\mu$ is $X_{J^\mu}=d-1$. 

In $d=2$, after switching to Euclidean coordinates $(x_1,x_2)=(x_1,ix_0)$, a scaling field $\Phi(x)$ is characterized by the dimensions $(\Delta_\Phi,\bar{\Delta}_\Phi$) which determine the scaling dimension $X_\Phi=\Delta_\Phi+\bar{\Delta}_\Phi$ and the Euclidean spin $s_\Phi=\Delta_\Phi-\bar{\Delta}_\Phi$. Equation (\ref{conservation}) can be rewritten in the form
\EQ
\partial_z\bar{J}=\partial_{\bar{z}} J\,,
\label{conservation2}
\EN
where $z=x_1+ix_2$, $\bar{z}=x_1-ix_2$, and $J$ and $\bar{J}$ are current components with dimensions
\EQ
\Delta_J=\bar{\Delta}_{\bar{J}}=1\,,\hspace{1.5cm}\bar{\Delta}_J=\Delta_{\bar{J}}=0\,. 
\label{dimensions}
\EN
The vanishing for both components of one of the dimensions yields $\partial_z\bar{J}=\partial_{\bar{z}} J=0$. It follows that a renormalization group fixed point possesses a conserved current if it contains fields $J$ and $\bar{J}$ with the dimensions (\ref{dimensions}). This is certainly the case in presence of a line of fixed points, since such a line is generated by a marginal scalar field $A$, namely a field with $\Delta_A=\bar{\Delta}_A=1$ which can be expressed as $A\sim J\bar{J}$. 

Hence, in $d=2$ a line of renormalization group fixed points implies the conservation equation (\ref{conservation}) and the invariance of the action under a continuous symmetry. It can happen, however, that the order fields of a statistical model renormalizing on that action do not carry a representation of the continuous symmetry, so that the relevant symmetry remains discrete and can be broken spontaneously at low temperature. Instead, no finite temperature transition arises when the order fields carry a representation of the continuous symmetry, since the latter cannot break spontaneously in $d=2$. It follows that in $d=2$ a line of renormalization group fixed points and the absence of a finite temperature transition to an ordered phase are sufficient conditions for continuous internal symmetry.

Lines of fixed points are uncommon. The physically relevant case that has been known is that of the two-dimensional Gaussian model (see \cite{DfMS} and, concisely, \cite{colloquium}) with action
\EQ
{\cal A}_G=\frac{1}{4\pi}\int d^2x\,(\nabla\varphi)^2\,,
\label{free}
\EN
in which the conserved current components $J\sim \partial_z\varphi$ and $\bar{J}\sim\partial_{\bar{z}}\varphi$ follow from the fact that $\varphi$ is dimensionless. The equation of motion $\partial_z\partial_{\bar{z}}\varphi=0$ allows the decomposition \EQ
\varphi(x)=\phi(z)+\bar{\phi}(\bar{z})\,,
\label{phi}
\EN
which is in turn consistent with the logarithmic correlator $\langle\varphi(x)\varphi(0)\rangle=-\ln|x|=-\frac{1}{2}(\ln z+\ln\bar{z})$. Besides $J$ and $\bar{J}$, proper scaling fields with power law correlators are the exponentials $e^{2i[p\phi(z)+\bar{p}\bar{\phi}(\bar{z})]}$ with dimensions $(\Delta,\bar{\Delta})=(p^2,\bar{p}^2)$. The energy density field $\varepsilon=\cos 2b\varphi$ has $\Delta_\varepsilon=\bar{\Delta}_\varepsilon=b^2$, with $b^2$ parameterizing the line of fixed points. 

The continuous symmetry of the model (\ref{free}) is a one-parameter shift symmetry which becomes $O(2)$ if the shift parameter is an angular variable. The basic $O(2)$-invariant lattice spin model, namely the XY ferromagnet (see e.g. \cite{Cardy_book}) with Hamiltonian
\EQ
H_{XY}=-J\sum_{\langle x,y\rangle}{\bf s}(x)\cdot{\bf s}(y)\,,\hspace{1cm}{\bf s}(x)=(\cos\theta(x),\sin\theta(x))\,,
\EN
where ${\bf s}(x)$ is the spin variable at site $x$ of a regular lattice and $\sum_{\langle x,y\rangle}$ is the sum over nearest neighbor pairs, indeed renormalizes at criticality in $d=2$ onto the Gaussian model (\ref{free}), with 
\EQ
\theta=\frac{1}{2b}(\phi-\bar{\phi})\,.
\EN
Continuous symmetries do not break spontaneously in $d=2$ \cite{MW,Hohenberg,Coleman_goldstone}, but the $XY$ model exhibits a different type of transition known as Berezinskii-Kosterlitz-Thouless (BKT) transition \cite{Berezinskii,KT}. Lowering the temperature in the XY model amounts to increasing $b^2$ in the Gaussian model. Below a critical temperature $T_c$ corresponding to $b^2=1$, the field $\varepsilon$ becomes irrelevant ($X_\varepsilon>2$), thus producing the BKT phase with $\langle{\bf s}(x)\rangle=0$ and infinite correlation length\footnote{The addition to (\ref{free}) of the $\varepsilon$ perturbation yields the sine-Gordon model, whose fundamental excitations are topological modes (solitons) interpolating between different minima of the cosine potential. For $b^2>1$, $\varepsilon$ is irrelevant and only topologically neutral excitations are allowed. Hence, the BKT transition occurring at $b^2=1$ is referred to as {\it topological}.}.

For ferromagnets, the tendency of the spins to point in the same direction makes irrelevant -- in  the renormalization group sense -- microscopic details such as the lattice structure (e.g., in two dimensions, square vs triangular). On the other hand, for antiferromagnets ($J<0$), and more generally for non-ferromagnets, the lattice structure matters also for critical properties. A particularly interesting example for the subject of this paper is that of the square lattice three-state Potts antiferromagnet,  whose Hamiltonian is invariant under the permutational group $\mathbb{S}_3$.  The order variable is the staggered magnetization $\Sigma(x)$, which is odd under exchange of even and odd sublattices. As a combined result of sublattice parity and $\mathbb{S}_3$-invariant Hamiltonian, the critical properties of the model are described\footnote{At criticality, this antiferromagnet is solved on the lattice \cite{Baxter_AF} and this allows the field theoretical identification.} by the Gaussian model (\ref{free}) with $\Sigma=e^{i\theta}=e^{i(\phi-\bar{\phi})/2b}$ and $b^2=3/4$ \cite{PottsAF1}.  This exact result shows that the infinitely many degenerate ground states of the antiferromagnet are labeled by the continuous angular variable $\theta$, and that the discrete symmetry of the lattice model is enhanced near criticality to the continuous $O(2)$ symmetry of the Gaussian model. Since now the temperature is not related to $b^2$, which is fixed, absence of ordered phases for continuous symmetries in $d=2$ implies that the critical point has to occur at $T=0$, and this is indeed the case \cite{Baxter_AF}. Other planar lattices splitting into two equal sublattices realize the same pattern of $T=0$ Gaussian criticality for other values of $b^2$ \cite{fpu}. For lattices without this property the symmetry remains $\mathbb{S}_3$ and the critical properties fall in the universality class of the ferromagnet \cite{PottsAF2}. These theoretical results are confirmed by numerical studies on several planar lattices \cite{q3lattice}. In $d>2$ the enhancement to $O(2)$ symmetry continues to hold on lattices possessing the sublattice exchange symmetry, in particular on the hypercubic lattice. This is confirmed by numerical studies of the $d=3$ model on the cubic lattice \cite{BGJ,WSK,GH}. Hence, due to the enhancement to continuous internal symmetry, the lower critical dimension of the three-state Potts antiferromagnet on the hypercubic lattice is\footnote{Note that $X_\varepsilon=2b^2=3/2$ on the square lattice rules out the assumption, sometimes invoked in the literature, that the correlation length critical exponent $\nu=1/(d-X_\varepsilon)$ is infinite at the lower critical dimension.} $d_L=2$. 

Also the two-dimensional Ashkin-Teller model (two coupled Ising models) renormalizes at criticality on the Gaussian action (\ref{free}) \cite{KB}. However, the order fields, which are the two Ising spin fields, cannot be themselves expressed in terms of the field (\ref{phi}) \cite{KB,DG_AT}. Hence, the symmetry relevant for the Ashkin-Teller model remains discrete and can be spontaneously broken also in $d=2$. This is why the model possesses an ordered phase \cite{Baxter}.

\section{Ising spin glass}
\label{Ising_glass}
\subsection{Generalities}
The random bond Ising model \cite{EA} is defined by the lattice Hamiltonian
\EQ
H=-\sum_{\langle x,y\rangle}J_{xy}\, \sigma(x)\sigma(y)\,,\hspace{1.5cm}\sigma(x)=\pm 1\,,
\label{Ising}
\EN
where the sum is taken over nearest neighbors. Quenched disorder and frustration are realized through random couplings $J_{xy}$ drawn from a probability distribution $P(J_{xy})$ allowing for both ferromagnetic and antiferromagnetic values, for example the bimodal distribution
\EQ
P(J_{xy})=p\,\delta(J_{xy}-1)+(1-p)\,\delta(J_{xy}+1)\,,
\label{bimodal}
\EN
in which $1-p$ is the fraction of antiferromagnetic bonds.  Ferromagnetic order corresponds to $[\langle\sigma(x)\rangle]\neq 0$,  
where we denote by $\langle\cdots\rangle$ the thermal average and by $[\cdots]$ the disorder average.  

When the fractions of ferromagnetic and antiferromagnetic bonds are comparable the system is magnetically disordered at all temperatures.  However, for low enough temperature, a spin glass phase is expected to correspond to correlations (memory effects) between states of the system separated by evolution over a time interval $t\to\infty$.  In the equilibrium framework,  this amounts to correlations (overlaps) among independent copies\footnote{These copies are also called {\it real} replicas to distinguish them from the $n\to 0$ auxiliary replicas introduced by (\ref{trick}) below.}  of the system with the same disorder realization $\{J_{xy}\}$,  namely to the Hamiltonian
\EQ
H_N=-\sum_{\langle x,y\rangle}J_{xy}\sum_{a=1}^N\sigma^{(a)}(x)\sigma^{(a)}(y)\,,\hspace{1.5cm}\sigma^{(a)}(x)=\pm 1\,,
\label{lattice}
\EN
where $a$ labels the copies.  Introducing the overlap variables \cite{EA,BY}
\EQ
q^{a,b}(x)=\sigma^{(a)}(x)\sigma^{(b)}(x)\,,\hspace{1cm}a\neq b\,,
\label{overlap}
\EN
the overlap order parameter
\EQ
[\langle q^{a,b}(x)\rangle]=[\langle\sigma^{(a)}(x)\rangle\langle\sigma^{(b)}(x)\rangle]
\label{glass_op}
\EN
is expected to characterize the spin glass transition.

\begin{figure}[t]
    \centering
    \begin{subfigure}[h]{0.35\textwidth}
        \includegraphics[width=\textwidth]{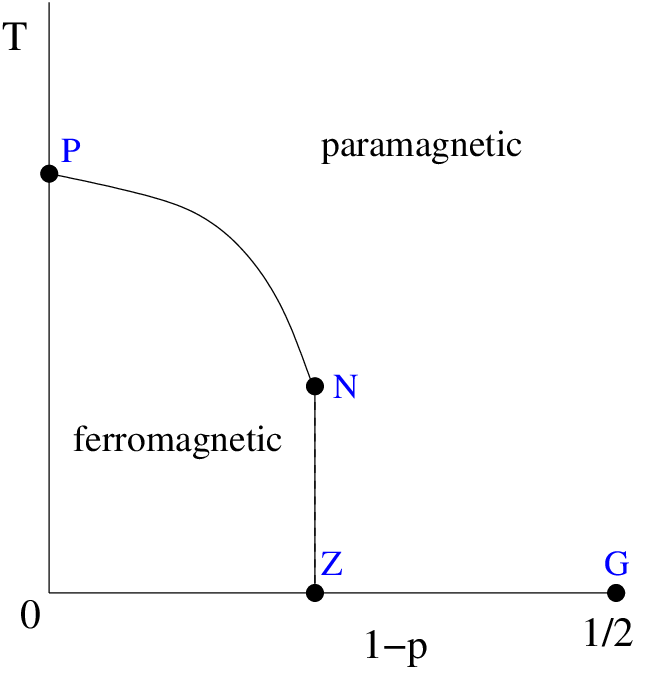}
    \end{subfigure}\hspace{3cm}%
    \begin{subfigure}[h]{0.35\textwidth}
        \includegraphics[width=\textwidth]{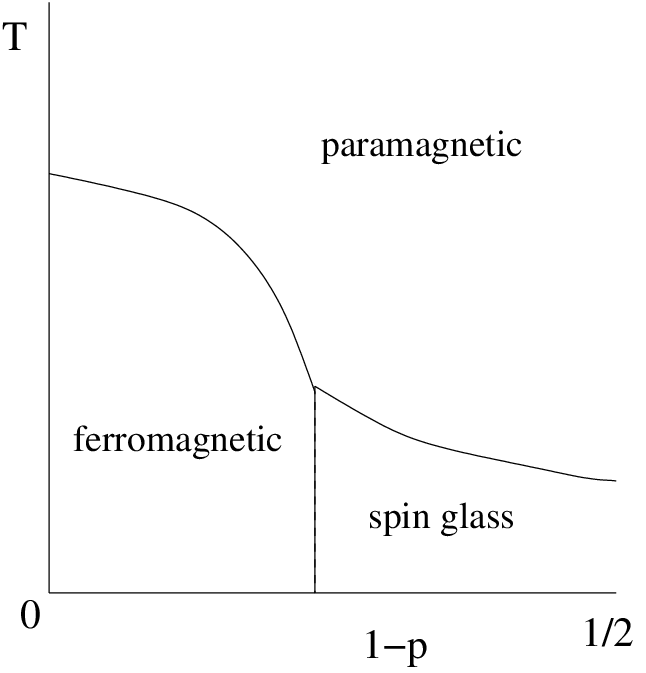}
    \end{subfigure}
    \caption{The phase diagram of the $\pm J$ random bond Ising model on the hypercubic lattice is symmetric under $p\to 1-p$.  \textbf{Left:} $d=2$ with the indication of the magnetic (P, N, Z) and spin glass (G) renormalization group fixed points.  \textbf{Right:} $d=3$.}
    \label{pd}
\end{figure}

The phase diagrams obtained from numerical simulations (see e.g. \cite{KR} and references therein) in two and three dimensions are shown in Figure~\ref{pd}. The absence of finite temperature transition to a spin glass phase in $d=2$ is known from simulations (since \cite{McMillan}, at least) and analytical extrapolations \cite{Nishimori_T=0} for bimodal and Gaussian disorder distributions on the square lattice. The spin glass transition in $d=3$ is detected by finite size scaling studies of $[\langle q^{a,b}(x)q^{a,b}(y)\rangle]$ and higher-point overlap correlation functions (see \cite{KKY,HPV_3D,B-J}). On the other hand, the nature of the transition and of the spin glass phase remained debated \cite{BY,KR,Dahlberg}.

The following observation is very useful from the theoretical point of view.  Since the average over quenched disorder is taken on the free energy $-\ln Z$, where 
\EQ
Z=\sum_{\{\sigma^{(1)}(x)\},\cdots,\{\sigma^{(N)}(x)\}}e^{-H_N/T}
\label{Z}
\EN
is the partition function with an assigned disorder configuration,  the identity 
\EQ
\ln Z=\lim_{n\to 0}\frac{Z^n-1}{n}\,
\label{trick}
\EN
shows that the effect of the disorder average is that of coupling $n\to 0$ replicas of the system with Hamiltonian (\ref{lattice}). Hence, we end up with $Nn$ Ising models coupled only by the interactions produced by the disorder average. This ``replica method" allows to use for disordered systems the results on symmetries, scaling and renormalization known for pure systems.

\subsection{$d=2$}
Recently the first exact results for spin glass criticality in finite dimension have been obtained in $d=2$ \cite{Ising_glass}. We start by recalling the steps of the derivation referring the reader to \cite{Ising_glass} for more details. 

In $d=2$ the limit $n\to 0$ involved in the replica method can be taken in an {\it exact} way at renormalization group fixed points \cite{random} (\cite{colloquium} for a review). At such points there is conformal invariance\footnote{Conformal invariance is ordinarily present at fixed points of lattice models with short range interactions and homogeneous and isotropic scaling limit (see e.g. \cite{DfMS,Cardy_book}). No obstructions to homogeneity and isotropy are present for the scaling limit of Ising spin glasses on regular lattices, and successful numerical tests of conformal invariance at the spin glass fixed point G have been presented in \cite{AHHM,BLdM}. We mention that for non-random Lagrangian field theories with several parameters and fixed symmetry the requirement of conformal invariance is expected to severely constrain the allowed values of the parameters, as a manifestation of the difficulty to find lines of conformally invariant points other than in the two-dimensional Gaussian model (see \cite{colloquium} for $O(N)$ symmetry); this is illustrated in \cite{RC} for an $O(2)$-invariant model of elasticity.}, which in $d=2$ has infinitely many generators \cite{DfMS} and yields infinitely many quantities to be conserved in the scattering framework available once one of the two spatial dimensions is considered as imaginary time. It follows that initial and final states are kinematically identical (complete elasticity). In addition, scale and relativistic invariance make scattering amplitudes energy-independent. The scattering problem then becomes exactly solvable and has provided the first exact results for random criticality \cite{random,DT2,DL_ON1,DL_ON2,DL_softening,random_line}, as well as progress on difficult problems without disorder \cite{DT1,ising_vector,DDL_nematic,potts_qr,RPN_universality}. With the notation $S_{\alpha\beta}^{\gamma\delta}={}_\alpha^\delta\times_\beta^\gamma$ for a generic scattering amplitude, the crossing and unitarity equations \cite{ELOP} take the simple form \cite{colloquium,paraf}
\begin{equation}
S_{\alpha\beta}^{\gamma\delta}=[S_{\alpha\delta}^{\gamma\beta}]^*\,,
\label{cross}
\end{equation} 
\begin{equation}
\sum_{\epsilon,\phi} S_{\alpha\beta}^{\epsilon\phi}[S_{\epsilon\phi}^{\gamma\delta}]^*=\delta_{\alpha\gamma}\delta_{\beta\delta}\,,
\label{unitarity}
\end{equation}
respectively.

\begin{figure}[t]
\centering
\includegraphics[width=14cm]{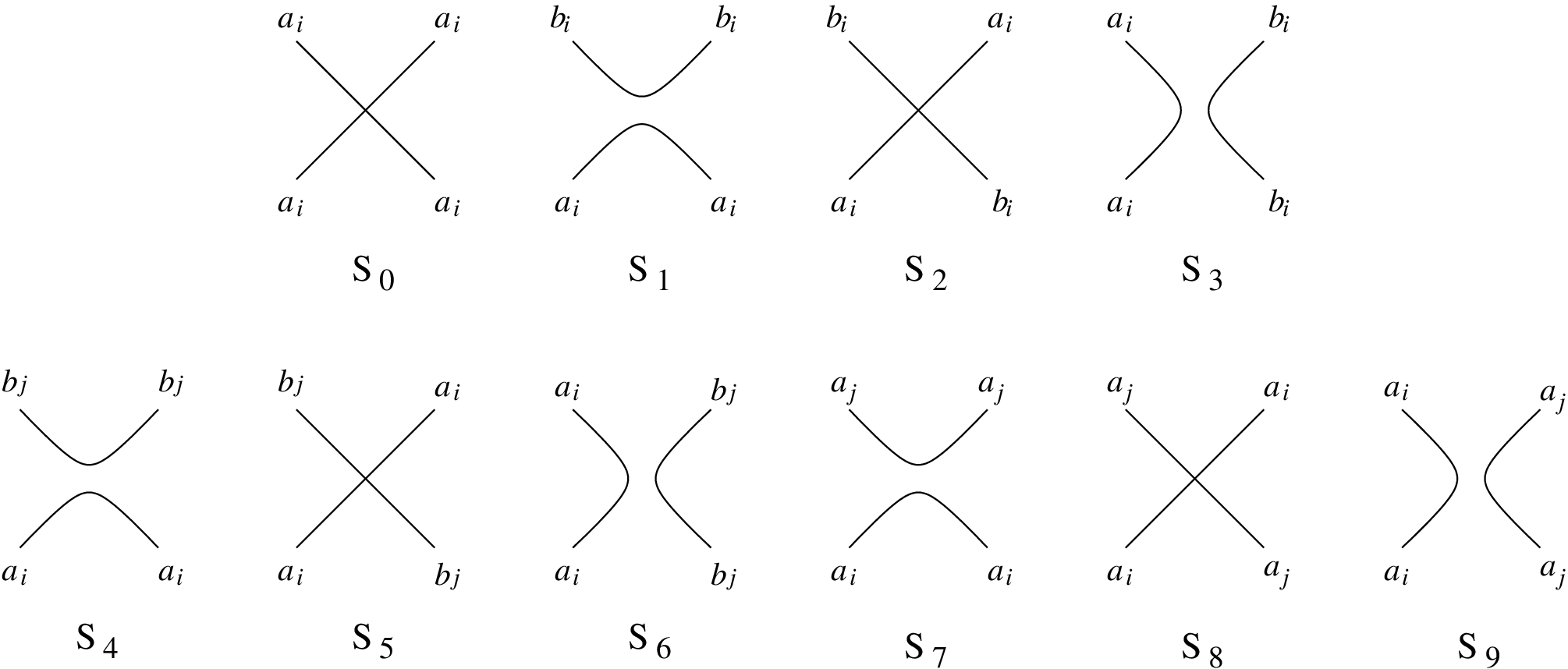}
\caption{Magnetic and overlap scattering amplitudes for the random bond Ising model. $a_i$ labels a particle excitation in the $i$-th replica of the $a$-th copy ($a \neq b$, $i \neq j$). Overlap amplitudes involve different copies. Time runs upwards.}
\label{amplitudes}
\end{figure}

The scattering particles are the fundamental collective excitation modes of the system. In the present case they are associated to the Ising spins $\sigma^{(a)}_i$ and can be denoted by $a_i$, where $a=1,2,\ldots,N$ is the copy index and $i=1,2,\ldots,n$ is the replica index. The elastic scattering amplitudes allowed by spin reversal symmetry $\sigma^{(a)}_i\to-\sigma^{(a)}_i$ and by permutational symmetry (of the $N$ copies as well as of the $n$ replicas) are shown in Figure~\ref{amplitudes}. The specialization of (\ref{cross}) to these amplitudes yields
\begin{align}
S_k&=S_k^*\,,\hspace{3.4cm} k=0,2,5,8
\label{crossing1}\\
S_k&=S_{k+2}^*\equiv X_k+iY_k\,,\hspace{1cm}k=1,4,7\,,
\label{crossing2}
\end{align}
with $X_k$ and $Y_k$ real, so that the unitarity\footnote{The unitarity of the scattering matrix should not be identified with the reflection positivity of the associated field theory, which in the literature is very often also called ``unitarity". The unitarity of the scattering matrix -- the only one we refer to -- expresses the fact that probabilities sum to one and always holds for field theories describing the scaling limit of statistical lattice models, even when these theories do not satisfy reflection positivity \cite{CM_yl}. This is the case, in particular, of the spin glass, for which reflection positivity is lost due to the $n\to 0$ limit involved in (\ref{trick}). See \cite{CM_yl,Zamo_saw,Zamo_perc} for the exact unitary scattering matrices of other well known statistical problems with non-reflection-positive scaling limit, respectively the Yang-Lee edge singularity, self-avoiding walks, and percolation. The latter two involve limits analogous to the present $n\to 0$ (see also \cite{BR} on this type of limits).} equations (\ref{unitarity}) take the form
\begin{align}
&S_0^2+(N-1)[X_1^2+Y_1^2+(n-1)(X_4^2+Y_4^2)]+(n-1)(X_7^2+Y_7^2)=1\,,
\label{uni1}\\
&2S_0 X_1+(N-2)[X_1^2+Y_1^2+(n-1)(X_4^2+Y_4^2)]+2(n-1)(X_4 X_7+Y_4 Y_7)=0\,,
\label{uni2}\\
&2 S_0 X_4+2(X_1X_7+Y_1Y_7)+(N-2)[2(X_1X_4+Y_1Y_4)+(n-2)(X_4^2+Y_4^2)]\nonumber\\
&+2(n-2)(X_4X_7+Y_4Y_7)=0\,,\label{uni3}\\
&2 S_0X_7+(n-2)(X_7^2+Y_7^2)+(N-1)[2(X_1X_4+Y_1Y_4)+(n-2)(X_4^2+Y_4^2)]=0,\\
&X_k S_{k+1}=0\,,\hspace{2.4cm}k=1,4,7\,,\label{uni9}\\
&X_k^2+Y_k^2+S_{k+1}^2=1\,,\hspace{1cm}k=1,4,7\,.
\label{uni10}
\end{align}

The solutions of these equations\footnote{Notice that $n$ appears as a parameter that does need to be an integer, so that the limit $n\to 0$ can be taken straightforwardly. The scattering framework presently provides the only exact access to critical quenched disorder \cite{random}, for which more traditional implementations of conformal invariance \cite{DfMS} remain impracticable (see the discussion in \cite{colloquium}).} are the renormalization group fixed points with the symmetries specified above. Some of these solutions, however,  are not of interest for the random bond Ising model.  Indeed,  the symmetries allow for the addition to the Hamiltonian (\ref{lattice}) of terms coupling the different copies, as in the $N$-color Ashkin-Teller model \cite{random_line}, while in our present case the $Nn$ Ising models are correlated only by the disorder average.  In the spin glass, the correlations (overlaps) among the copies cannot depend on their number $N$, which is not a physical parameter of the problem.  Hence,  the exact fixed point equations (\ref{uni1})-(\ref{uni10}) should admit solutions which, in the limit of interest $n=0$, are $N$-independent.  Remarkably,  at $n=0$ the $N$-dependent terms in the equations cancel when
\EQ
S_1=S_4\,,
\label{s4}
\EN
thus providing the solutions of our interest.  These solutions have been discussed in \cite{Ising_glass} and split into two subsets depending on whether the amplitudes $S_1=S_4$ vanish or not.  In the first case the copies are uncorrelated (no overlap) and we are in the magnetic sector which can be studied already with $N=1$.  The solutions in this sector correspond to the magnetic fixed points (see  \cite{McMillan,MC,PHP,HPtPV} for numerical studies) of the phase diagram of Figure~\ref{pd},  namely the fixed point P of the pure system, at which disorder is marginally irrelevant \cite{DD_81,DD_83,Shalaev,Ludwig}, the Nishimori multicritical point N \cite{Nishimori,Nishimori_book,LdH1,GD_Nishimori}, and the zero temperature fixed point Z.  The fixed point G of Figure~\ref{pd}, which rules the spin glass critical behavior,  can correspond to two solutions with nonzero overlap\footnote{The amplitudes not indicated in (\ref{random_line}) and (\ref{no_en}) follow from (\ref{crossing2}), (\ref{uni10}) and (\ref{s4}); $(\pm)$ indicates that both signs are allowed.} \cite{Ising_glass},
\EQ
S_0=1\,,\hspace{.5cm}X_1=S_7=0\,,\hspace{.5cm}Y_1\in[-1,1]\,,
\label{random_line}
\EN
and
\EQ
S_0=\sqrt{2}\,,\hspace{.5cm}X_1=\pm Y_1=(\pm)\frac{1}{\sqrt{2}}\,,\hspace{.5cm}S_7=\frac{1}{\sqrt{2}}\,(1\pm i)\,.
\label{no_en}
\EN

As already observed for the Potts antiferromagnet, critical behavior in non-ferromagnetic transitions depends on the microscopic realization, which for the random bond Ising model is specified by the lattice structure and the disorder distribution. The spin glass fixed point G is far from the ferromagnetic phase, and one can expect the existence of microscopic realizations renormalizing on (27) as well as of microscopic realizations renormalizing on (28). What we have to do is to compare the results obtained in the continuum with the available lattice results. Very remarkably, solution (\ref{random_line}) has $Y_1$ taking values in an interval and corresponds to a line of fixed points. As observed in \cite{Ising_glass}, this finding immediately explains the following features emerging from the existing numerical studies of the Ising spin glass in $d=2$. On one side, the issue of universality at the fixed point G has been discussed in the numerical literature \cite{KLC,Hartmann,PtPV,Fernandez1,LC,KW}. Several disorder distributions on the square lattice were compared in \cite{LC}, where nonuniversality was clearly observed and it was concluded that the critical exponents can vary continuously upon tuning an extra parameter in some disorder distributions. Special accuracy turned out to be needed for the comparison of the two microscopic realizations which are most often investigated, namely the Gaussian and $\pm J$ distributions on the square lattice, for which the extensive numerical study of \cite{KW} obtained the values $1.27319(9)$ and $1.279(2)$, respectively, for the domain wall fractal dimension. The existence of solution (\ref{random_line}) finally explains these numerical results since it allows different microscopic realizations to renormalize on distinct fixed points on a line. The fact that these points can be arbitrarily close accounts for small differences as the one observed in \cite{KW}, and allows for exponents which vary continuously as a disorder distribution is deformed as concluded in \cite{LC}. On another side, the fact that (\ref{random_line}) is a free boson (pure transmission with amplitude $+1$) for $Y_1=0$,  and almost a free boson for $|Y_1|\ll 1$,  accounts for the numerical conclusion of \cite{Fernandez2} that the square lattice model with $\pm J$ disorder behaves ``almost but not quite" as a free boson. Solution (\ref{no_en}), instead, is far from free bosonic behavior.

The sufficient conditions for continuous internal symmetry of section~\ref{symmetry} now allow us to add an important observation about solution (\ref{random_line}). Indeed,  the presence of a line of fixed points and the absence of finite temperature transition to a spin glass phase are the signature of enhancement to continuous symmetry.  At the same time,  the absence of a finite temperature transition observed in the simulations and the analytical extrapolations\footnote{The analytical extrapolations are performed for disorder distributions possessing a Nishimori line on planar self-dual lattices \cite{Nishimori_T=0}.} finally finds a natural explanation in the fact that continuous symmetries do not break spontaneously in $d=2$.  The fact that the line of fixed points is parameterized by the overlap amplitude $Y_1$ shows the essential role played by the correlations among the copies.  Similarly to what we saw in section~\ref{symmetry} for the Potts antiferromagnet,  the symmetry enhancement can be traced back to a complexity of the ground state pattern,  produced in the present case by both frustration and disorder.  The role of disorder in the symmetry enhancement is shown by the fact that the line of fixed points arises only at $n=0$.  

The continuous internal symmetry is a one-parameter shift symmetry. $N$-independence implies that the spin glass order parameter $[\langle q^{a,b}\rangle]$ does not depend on the pair of different copies $a$ and $b$. Hence, we have a one-component order parameter,
\EQ
[\langle q^{a,b}\rangle]=q\,,
\label{q}
\EN
and the continuous shift variable -- let us call it $s$ -- is not an angular variable.  It is worth stressing that the parameter $Y_1$ of solution (\ref{random_line}) and the shift variable $s$ are independent: $Y_1$ labels the critical behavior of different microscopic realizations allowing for enhancement to continuous internal symmetry; $s$ labels infinitely many ground states for each value of $Y_1$.  In the analogy with the Potts antiferromagnet they correspond to $b^2$ and $\theta$,  respectively. At variance with the Potts antiferromagnet, however, $s$ is not an angular variable\footnote{More generally, in the comparison with the Potts antiferromagnet, we see that the mechanism producing the enhancement to continuous symmetry is much more subtle for the spin glass. One starts again from discrete symmetries -- spin reversal and permutations of the $Nn$ Ising models -- but the line of fixed points and symmetry enhancement arise only at $n=0$.}.

The absence of finite temperature glass transition observed in the simulations of two-dimensional Ising spin glasses had somehow remained puzzling from the point of view of the general theory of critical phenomena and, more specifically, of the notion of lower critical dimension $d_L$, which is the spatial dimensionality above which a phase transition starts to occur. Indeed, since the Ising spin glass Hamiltonian has discrete internal symmetry and $d_L=1$ for discrete symmetries \cite{Cardy_book}, one would at first expect a finite temperature transition in $d=2$. The fact that it is not observed is now immediately explained by our finding of enhancement to continuous symmetry, since $d_L=2$ for continuous symmetries. Within the droplet picture -- one of the pictures proposed for the Ising spin glass (see e.g. \cite{KR} for a review) -- the presence of the transition is characterized through the sign -- to be determined in the numerical simulations -- of a droplet exponent $\vartheta$. Clearly, our exact result of symmetry enhancement is independent of such characterization based on $\vartheta$ and does not contradict it. In a similar way, the exponent $\vartheta$ does not arise in the study of \cite{Nishimori_T=0} about the absence of finite temperature transition in $d=2$. More generally, concerning critical exponents of two-dimensional Ising spin glasses, the nonuniversality discussed above in this subsection implies that they cannot be determined in field theory, which is formulated in the continuum: the dependence of critical exponents on the microscopic realization -- disorder distribution and lattice structure: square, triangular, etc. -- implies that they can only be identified through {\it case by case lattice studies}. In the context of magnetism, this ``case by case problem" is general and well known for non-ferromagnetic transitions: only for ferromagnetic transitions lattice details are irrelevant and exponents are universal \cite{Cardy_book}. In particular the numerical determination of $\vartheta$ for a microscopic realization implies nothing about the sign of this exponent for a different two-dimensional realization. It is then remarkable that field theory -- establishing the presence of a line of fixed points -- now provides the first and unifying explanation of more structural properties observed in the numerical studies of Ising spin glass criticality in $d=2$, namely the presence and specific features of nonuniversality, the absence of finite temperature transition, and the almost-free-bosonic behavior of \cite{Fernandez2}. 

The fact that some microscopic realizations (in particular the bimodal and Gaussian disorder distributions on the square lattice) do not exhibit a finite temperature transition to a spin glass phase does not imply that such a transition can never occur in $d=2$.  There may be microscopic realizations for which the symmetry remains discrete and can be spontaneously broken at low temperature.  In such a case,  solution (\ref{no_en}) would provide the fixed point ruling critical behavior along the transition line in the temperature-disorder plane.  Depending on sign choices,  (\ref{no_en}) yields a small discrete set of fixed points: absence of enhancement to continuous symmetry may still allow classes of microscopic realizations with different critical behavior.

\subsection{$d>2$}
The internal symmetry characterizes the transition properties in all dimensions, without qualitative differences above the lower critical dimension. We deduced from exact results in $d=2$ that the Ising spin glass possesses a continuous one-parameter internal symmetry for microscopic realizations which include the bimodal and Gaussian disorder distributions on the square lattice.  The enhancement to continuous symmetry will occur also for microscopic realizations in $d>2$.  Bimodal and Gaussian disorder distributions on the hypercubic lattice,  which reduces to the square lattice in $d=2$,  are expected to be among such realizations. We now know that the lower critical dimension is that of continuous symmetries, namely $d_L=2$. The key difference in $d>2$ is that the continuous symmetry can break spontaneously allowing a finite temperature transition to a spin glass phase. For fixed values of temperature and disorder strength inside this phase, the overlap order parameter (\ref{q}) takes continuous values in an interval $[-Q,Q]$ as the shift variable $s$ changes; $Q\to 0$ as the transition to the paramagnetic phase is approached. The Goldstone boson associated to the spontaneous breaking of the continuous shift symmetry provides the lowest-energy excitation modes of the spin glass phase.

Our exact result that the continuous shift symmetry only arises at $n=0$ implies a difficulty for finding a Landau-Ginzburg description of the spin glass transition, and then a difficulty for determining the upper critical dimension $d_c$ (i.e. the dimension below which the critical exponents differ from mean field values). It can be observed that the recent determination of exact critical exponents \cite{GD_Nishimori} yields $d_c=6$ at the Nishimori point, which for $d>2$ is the multicritical point where the ferromagnetic, paramagnetic and spin glass phases meet \cite{Nishimori,Nishimori_book,LdH1} (Figure~\ref{pd}).

The mean field solution \cite{Parisi} of the model with infinite-range interactions, which corresponds to $d=\infty$, is obtained for $N=1$, a value for which the real overlaps (\ref{overlap}) are not defined and not measurable. On the other hand, the mean field procedure mathematically leads to consider the overlaps $[\langle q^{i,j}\rangle]$ between auxiliary replicas rather than among real replicas, and the characterizing property of $[\langle q^{i,j}\rangle]$ in the mean field solution is that it takes continuous values in the spin glass phase \cite{Parisi}. The fact that this is the same property that $[\langle q^{a,b}\rangle]$ acquires from the spontaneous breaking of the continuous symmetry is not likely to be a coincidence. Indeed, we saw that the symmetry enhancement is a property of $Nn\to 0$ Ising models. It is probably the $N$-independence at $n=0$ which allows the mean field $[\langle q^{i,j}\rangle]$ to inherit properties of $[\langle q^{a,b}\rangle]$\,\footnote{If future lattice studies will identify microscopic realizations with finite temperature spin glass transition in $d=2$, these will correspond to solution (28) and we will know that in $d\geq 2$ there are microscopic realizations for which the enhancement to continuous symmetry does not occur. For such microscopic realizations, the above mechanism yielding an order parameter $[\langle q^{a,b}\rangle]$ with continuous support in $d>2$ will not apply. The scenario of an order parameter with discrete support for some microscopic realizations is compatible with the $d=\infty$ mean field solution of \cite{Parisi}. Indeed, it is not known if this is the only mean field solution and if it can apply to all microscopic realizations.}.

It is worth recalling that spin glass mean field theory in $d=\infty$ has been at the origin of concerns about the use of the replica method. Indeed, the original mean field solution of \cite{SK} involved an unphysical negative entropy for the spin glass phase. While this problem was initially attributed to the limit $n\to 0$ of (\ref{trick}), it was later shown in \cite{DaT} that the real issue was in the instability of the mean field solution with the auxiliary replicas treated symmetrically. A solution without negative entropy was then obtained in \cite{Parisi} breaking in a specific way the symmetry among those replicas inside the spin glass phase, where the spin glass order parameter $q$ is nonzero. We have used the method of auxiliary replicas at the spin glass fixed point G in $d=2$. At such point $q$ vanishes and -- even in the hypothesis of an analogy between $d=2$ and mean field theory in $d=\infty$ -- there is no issue of replica symmetry breaking\footnote{Our use of the limit $n\to 0$ at criticality is then analogous to that of the limit $Q\to 1$ which yields percolation within the $Q$-state Potts model \cite{Wu}. The exact field theoretical results for the crossing probabilities \cite{Cardy_crossing} and the three-point connectivity \cite{DV_3point} for critical percolation -- obtained in the $Q\to 1$ limit -- became mathematical theorems in \cite{Smirnov,ACSW}, respectively.}. 

Hence, it is clear that our results are technically uncorrelated to the replica symmetry breaking of \cite{Parisi}. What happens, as we saw, is that we show that the Ising spin glass allows for enhancement to a continuous symmetry whose spontaneous breaking in $d>2$ yields a spin glass order parameter with continuous support. The way we show this is new, exact and independent of what has been proposed in the past. We argued that the circumstance that also the mean field spin glass order parameter of \cite{Parisi} has continuous support -- in spite of the fact that it does not involve the real replicas -- may not be a coincidence but, clearly, we did not prove that the mean field mechanism of \cite{Parisi} extends to finite dimension. Conversely, no such proof is required for the validity of our results. The fact that the spin glass phase in $d>2$ corresponds to the spontaneous breaking of a continuous symmetry is not taken into account by any of the scenarios proposed in the past (see e.g. \cite{KR} for a review), since the enhancement to continuous symmetry was not known.

\section{Conclusion}
Recently, the first exact access to the spin glass regime in finite dimension has been obtained in $d=2$ \cite{Ising_glass}. This was achieved exploiting conformal invariance in the scattering framework to implement the replica method in an exact way and to obtain the renormalization group fixed points of the Ising spin glass. The problem involves $N$ copies (real replicas) of the system whose correlations (overlaps) characterize spin glass physics and are measured in the simulations, as well as $n$ auxiliary replicas (for each copy) needed to implement quenched disorder within the replica method. Very remarkably, the exact solution revealed the mechanism through which the necessary $N$-independence emerges precisely in the limit $n\to 0$ implied by the replica method. The presence of a line of fixed points, while unexpected because rarely realized in physically relevant contexts, immediately explains properties of Ising spin glass criticality observed in numerical studies in $d=2$, namely the presence and specific features of nonuniversality, and the almost-free-bosonic behavior of \cite{Fernandez2}. 

In this paper we observed that the existence of a line of fixed points in $d=2$ reveals the enhancement to a one-generator continuous internal symmetry (shift symmetry). We noted similarities and differences with the enhancement from discrete to continuous symmetry known for some antiferromagnets with infinitely many degenerate ground states. While the line of fixed points implies a condition of exact marginality which does not persist in $d>2$, the continuous symmetry that it reveals characterizes the model in all dimensions. In $d=2$ the symmetry enhancement immediately explains the absence of a finite temperature transition to a spin glass phase observed in the numerical simulations. In $d>2$,  instead, the spontaneous breaking of the continuous symmetry is allowed and yields a spin glass phase with an overlap order parameter which, for fixed temperature and disorder strength, takes continuous values in an interval. An order parameter with such a feature characterizes Parisi mean field solution of the model with infinite-range interactions, which corresponds to $d=\infty$. We argued that, in a subtle way, $N$-independence at $n=0$ may allow the mean field order parameter -- which involves only the auxiliary replicas -- to inherit the characteristic feature of the order parameter for the real replicas. 

The fact that the enhancement to continuous internal symmetry only arises at $n=0$ explains the difficulty to find a Landau-Ginzburg description of the spin glass transition and fixes its exceptional place within the theory of critical phenomena.

\end{document}